# New System for Secure Cover File of Hidden Data in the Image Page within Executable File Using Statistical Steganography Techniques

Md. Rafiqul Islam, A.W. Naji, A.A.Zaidan* and B.B.Zaidan

Department of Electrical and Computer Engineering, Faculty of Engineering,
International Islamic University Malaysia (IIUM), P.O. Box 10, 50728 Kuala Lumpur, Malaysia

**ABSTRACT—** A Previously traditional methods were sufficient to protect the information, since it is simplicity in the past does not need complicated methods but with the progress of information technology, it become easy to attack systems, and detection of encryption methods became necessary to find ways parallel with the differing methods used by hackers, so the embedding methods could be under surveillance from system managers in an organization that requires the high level of security. This fact requires researches on new hiding methods and cover objects which hidden information is embedded in. It is the result from the researches to embed information in executable files, but when will use the executable file for cover they have many challenges must be taken into consideration which is any changes made to the file will be firstly detected by untie viruses, secondly the functionality of the file is not still functioning. In this paper, a new information hiding system is presented. The aim of the proposed system is to hide information (data file) within image page of execution file (EXEfile) to make sure changes made to the file will not be detected by universe and the functionality of the exe.file is still functioning after hiding process. Meanwhile, since the cover file might be used to identify hiding information, the proposed system considers overcoming this dilemma by using the execution file as a cover file.

*(keyword):* Information Hiding, portable executable file, Steganography, Statistical Technique.

## I. INTRODUCTION

Information hiding is a general term encompassing many sub disciplines, is a term around a wide range of problems beyond that of embedding message in content. The term hiding here can refer to either making the information undetectable or keeping the existence of the information secret. Information hiding is a technique of hiding secret using redundant cover data such as images, audios, movies, documents, etc. This technique has recently become important in a number of application areas. For example, digital video, audio, and images are increasingly embedded with imperceptible marks, which may contain hidden signatures or watermarks that help to prevent unauthorized copy [1].It is a performance that inserts secret messages into a cover file, so that the existence of the messages is not apparent. Research in information hiding has tremendous increased during the past decade with commercial interests driving the field [1].

## II. PORTABLE EXECUTABLE FILE (PE-FILE)

The Program Loader that is a subset of the Windows System assumes the loading executable files into a virtual memory, so the executable files have the format that the Program Loader can identify, and the format is called PE (Portable Executable). It is necessary to know the PE format and RVA which is an address type used in the PE in order to understand the new methods for hiding information in the PE, so we briefly describe the format[2],[3].

And the address type.The planned system uses a portable executable file as a cover to embed an executable program as an example for the planned system.This section is divided into four parts [4]:

- Characteristics of executable files.
- Techniques Related with PE-File.
- Executable files types.
- PE File Layout

### A. Characteristics of Executable Files

The characteristics of the Executable file does not have a standard size, like other files, for example the image file (BMP) the size of this file is between (2-10 MB), Other example is the text file (TEXT) the size often is less than 2 MB.Through our study the characteristics of files have been used as a cover, it found that lacks sufficient size to serve as a cover for information to be hidden. For these features of the Executable file, it has unspecified size; it can be 650 MB like window setup File or 12 MB such as installation file of multi-media players. For taking advantage of this feature (disparity size) make it a suitable environment for concealing information without detect the file from attacker and discover hidden information in this file[3].

### B. Techniques Related with PE-File.

- RVA
  RVA is a position unit in the PE, and the RVA is used as an offset from the start-address of a PE file loaded on the memory. The start-address of a file on the memory is in Image Base that is one of the attributes of the PE file. For instance, if the Image Base of a file is 0x00400000 and one position of the file is 0x1000(RVA), the position on the memory will be 0x00401000[3],[4].

- PE Format

The header of PE format starts with MS-DOS stub that is used for printing a message, "This program cannot be run in DOS mode", if the operating system can't identify the PE on execution time.IMAGE_NT_HEADER located in the position after the MS-DOS stub has the information for the execution of a file, and consists of IMAGE_FILE_HEADER and IMAGE_OPTIONAL_HEADER. The IMAGE_FILE_HEADER has the information on the file, such as create time and machine type. The IMAGE_OPTIONAL_HEADER has the information on functions used in the file and on the start-address of the file on a memory, and the infor-mation is managed by IMAGE_DATA_DIRECTORY. A PE file except the header is composed of several sections that are basic unit of code or data within a PE or COFF file. IMAGE_SECTION_HEADER that is located in the position following to IMAGE_ OPTIONAL_HEADER has the information on each section. The information consists of PointerToRawData, SizeOfRawData, VirtualAddress, and VirtualSize. The PointerToRawData and the SizeOfRawData respectively mean the position of each section and the size of each section on the file. The VirtualAddress and the VirtualSize respectively mean the position of each section and the size of each section on the memory. The size of each section on the file is a multiple of FileAlignment that is in IMAGE_OPTIONAL_HEADER. If the amount of the data of a section is smaller than the size of the section that is allotted on compile time, the slack space of the section occurs. The common sections used in the PE include a .text that has program binaries, .data, .idata that has information on export and import functions, .edata, and .rsrc section. An .idata section has the information on import functions used in executable files during the period of an execution [5],[6].

C. *Executable File Types*

The number of different executable file types is as many and varied as the number of different image and sound file formats. Every operating system seems to have several executable file types unique to it. These types are [4],[5],[6]:

- EXE (DOS"MZ")

DOS-MZ was introduced with MS-DOS (not DOS v1 though) as a companion to the simplified DOS COM file format. DOS-MZ was designed to be run in real mode and having a relocation table of SEGMENT: OFFSET pairing. A very simple format that can be run at any offset, it does not distinguish between TEXT, DATA and BSS.The maximum file size of (code + data + bss) is one-mega bytes in size. Operating systems that use are: DOS, Win*, Linux DOS.

- EXE (win 3.xx "NE"):

The WIN-NE executable formatted designed for windows 3.x is the "NE" new-executable. Again, a 16-bit format, it alleviates the maximum size restrictions that the DOZ-MZ has. Operating system that uses it is: windows 3.xx.

- EXE (OS/2 "LE"):

The "LE" linear executable format was designed for IBM's OS/2 operating system by Microsoft supporting both 16 and 32-bit segments operating systems that are used in: OS/2, DOS.

- EXE (win 9x/NT "PE"):

With windows 95/NT a new executable file type is required, thus was born the "PE" portable executable. Unlike its predecessors, the WIN-PE is a true 32-bit file format, supporting releasable code. It does distinguish between TEXT, DATA, and BSS. It is in fact, a bastardized version of the common object file format (COFF) format. Operating systems that use it are: windows 95/98/NT/2000/ME/CE/XP.

- ELF:

The ELF, Executable Linkable Format was designed by SUN for use in their UNIX clone. A very versatile file format, it was later picked up by many other operating systems for use as both executable files and as shared library files. It does distinguish between TEXT, DATA and BSS.
TEXT: the actual executable code area.
DATA: "initialized" data, (Global Variables).
BSS : "un- initialized" data, (Local Variables).

D. *PE File Layout*

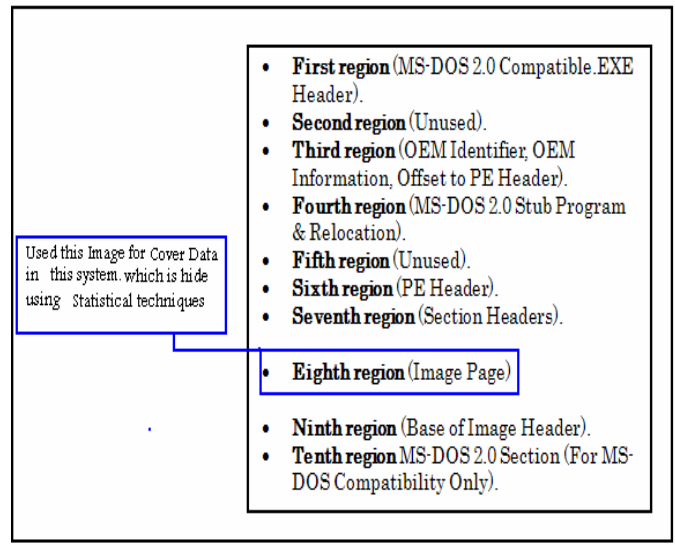

Figure 1.Typical 32-bit Portable .EXE File Layout.

III. STEGANOGRAPHY

Steganography is the art and science of writing hidden messages in such a way that no-one can realizes there is a hidden message in data (e.g. images file, documents file, sounds file … etc) except the sender and intended recipient.The word steganography is of Greek origin and means "covered, or hidden writing" .Cryptography obscures the meaning of a message, but it does not conceal the fact that there is a message, cryptography is the art of secret writing, which is intended to make a message unreadable by a third party but does not hide the message in a communication [1],[5]. Although steganography is separate and distinct from cryptography, but there are many analogies between the two, and some authors categorize steganography as a form of cryptography since hidden communication is a form of secret writing [2],[6]. A Watermark is a recognizable image or pattern in paper that appears as various shades of lightness/darkness when viewed by transmitted light (or when viewed by reflected light, atop a dark background), caused by thickness variations in the paper ,Digital Watermarking is the process of embedding information into a digital signal. The signal may be audio, pictures or video, for example. If the signal is copied, then the information is also carried in the copy. Steganography and digital watermarking are the same but the purpose of last one is use to copyright protection systems, which are intended to prevent or deter unauthorized copying of digital media. Steganography is an application of digital watermarking, where two parties communicate a secret message embedded in the digital signal [2].Classical Steganography often used methods of completely obscuring the message so it was unnoticeable to those who didn't know the specific covert method it was using for example Invisible Inks, which was able to write a confidential letter with any other non-value-confidential and usually write between lines. Modern Steganography refers to hide information in digital picture, audio or text files …etc , each one of this digitals data has a many techniques can use with it , for example in digital image the JPHide/JPSeek uses the coefficients in a JPEG to hide information, this method alter the image. In digital audio file several packages also exist for hiding data in audio files, Such as MP3Stego not only effectively hides arbitrary information, but also claims to be a partly robust method of watermarking MP3 audio files [4]. The Windows Wave format lets users hide data using Steghide, it alters the least significant bits (LSB) of data in the carrier medium [3].In a growing number of applications like digital rights management, covert communications, hiding executables for access control, annotation etc .all these application scenarios given the multimedia steganography techniques have to satisfy two basic requirements. The first requirement is perceptual transparency or noticeable perceptual distortion, i.e. cover object (object not containing any additional data) and stego object (object containing secret message) must be perceptually indiscernible [3]. The second constraint is high data rate of the embedded data. All the stego-applications, besides requiring a high bit rate of the embedded data, have need of algorithms that detect and decode hidden bits without access to the original multimedia sequence (blind detection algorithm) [4].

A. *Characterization of Steganography Systems*

Steganographic techniques embed a message inside a cover. Various features characterize the strength and weaknesses of the methods. The relative importance of each feature depends on the application [7].

- Capacity
  The notion of capacity in data hiding indicates the total number of bits hidden and successfully recovered by the Stego system.

- Robustness
  Robustness refers to the ability of the embedded data to remain intact if the stego-system undergoes transformation, such as linear and non-linear filtering; addition of random noise; and scaling, rotation, and loose compression.

- Undetectable
  The embedded algorithm is undetectable if the image with the embedded message is consistent with a model of the source from which images are drawn. For example, if a Steganography method uses the noise component of digital images to embed a secret message, it should do so while not making statistical changes to the noise in the carrier.Undetectability is directly affected by the size of the secret message and the format of the content of the cover image.

- Invisibility (Perceptual Transparency)
  This concept is based on the properties of the human visual system or the human audio system. The embedded information is imperceptible if an average human subject is unable to distinguish between carriers that do contain hidden information and those that do not. (Ross, 2005) It is important that the embedding occurs without a significant degradation or loss of perceptual quality of the cover.

- Security
  It is said that the embedded algorithm is secure if the embedded information is not subject to removal after being discovered by the attacker and it depends on the total information about the embedded algorithm and secret key.

B. *Statistical Steganography Techniques*

Statistical steganography techniques utilize the existence of "1-bits" Steganography  schemes, which embed one bit of information in a digital carrier. This is done by modifying the cover in such a way that some statistical characteristics change significantly if a "1" is transmitted. Otherwise, the cover is left UN changed. So the receiver must be able to distinguish unmodified covers from modified ones. A cover is divided into l (m) disjoint blocks B1...B l (m)[8]. A secret bit, mi is inserted into the ith block by placing "1" in to Bi if mi=1.Otherwise, the block is not changed in the embedding process.The detection of a specific bit is done via a test function which distinguishes modified block from unmodified block (1)[1],[8]:

$$f(B_i) = \begin{cases} 1 & \text{block } B_i \text{ was modified in the embedding process} \\ 0 & \text{otherwise} \end{cases}$$

... ... ... (1)

The function f can be interpreted as a hypothesis-test function and the test of null-hypothesis "block Bi was not modified "against the alternative hypothesis "block Bi was modified." Therefore, the whole class of such steganography systems statistical steganography .the receiver successively applies f to alaa cover-block Bi in order to restore every bit of the secret message. The main question which remains to be solved is how such a function f in (8) can be constructed. If they interpret f as a hypothesis-testing function , they can use the theory of hypothesis testing from mathematical statistics .Let us assume could find a formula h(Bi), which depends on some elements of the cover-block Bi, and knew the distribution of h(Bi), in the unmodified block (i.e,the Hypothesis holds in this case) could then use standard procedure to test if h(Bi), equals or exceeds a specific value. If managed to alter h(Bi) in the embedding process in a way that its expected value is 0 if the block Bi was not modified, and expected value is much greater otherwise , could test whether h(Bi) equals zero under the given distribution of h(Bi). Statistical steganography techniques are, however, difficult to apply in many cases. First, a good test statistic h(Bi) must be found which allows distinction between modified and unmodified cover-blocks. Additionally, the distribution of h (Bi) must be known for a "normal" cover; in most cases, this is quite a difficult task. In practical implementations many (quite questionable) assumptions are made in order to determine a closed formula for this distribution. As an example, wanted to construct a statistical steganography algorithem out of pitas' watermarking system, which is similar the patchwork approach of bender et al. Suppose every cover-block Bi is a rectangular set of pixels p(i)n,m .Furthermore, let S={s(i)n,m} be a rectangular pseudorandom binary pattern of equal size, where the number of one is S equals the number of zeros. Would assume that both the sender and receiver have access to S, which represents the stego-key in this application. The sender first splits the image block Bi into two sets, Ci and Di of equal size (i.e., putting all pixels with indices (n,m)into set C where the corresponding key bit n,m equals zero)[1],[8]:

$$C_i = \{p_{n,m}^{(i)} \in B_i | s_{n,m} = 1\}$$
$$D_i = \{p_{n,m}^{(i)} \in B_i | s_{n,m} = 0\}$$

... ... (2)

The sender then adds a value k > 0 to all pixels in the subset Ci but leaves all pixels in Di unchanged. In the last step, Ci and Di are merged to form the marked image block Bi. In order to extract the mark, the receiver reconstructs the sets Ci and Di. If the block contains a mark, all value in Ci will be larger than the corresponding values in the embedding step; thus testing the difference of the means of sets Ci and Di. If assumed that all pixels in both Ci and Di are independent identically distributed random variables with an arbitrary distribution, the test statistic [1],[8]:

$$q_i = \frac{\overline{C_i} - \overline{D_i}}{\hat{\sigma}_i}$$

with

$$\hat{\sigma}_i = \sqrt{\frac{\text{Var}[C_i] + \text{Var}[D_i]}{|S|/2}}$$

… … (3)

Where $\overline{C_i}$ denotes the mean over all pixels in the set Ci and Var [Ci] the estimated variance of the random variables in Ci , will follow a N(0,1) normal distribution asymptotically due to the central limit theorem . if a mark is embedded in the image block Bi , the expected value of q will be greater than zero. The receiver is thus able to reconstruct the ith secret message bit by testing whether the statistic qi of block Bi equals zero under the N (0, 1) distribution[7],[8].

## IV. METHODOLOGY

*A. System Overview*

The most important reason behind the idea of this system is that the programmers always need to create a back door for all of their developed applications, as a solution to many problems such that forgetting the password. This idea leads the customers to feel that all programmers have the ability to hack their system any time. At the end of this discussion all customers always are used to employ trusted programmers to build their own application. Programmers want their application to be safe anywhere without the need to build ethic relations with their customers. In this system a solution is suggested for this problem. The solution is to hide the password in the executable file of the same system and then other application to be retracted by the customer himself. Steganography needs to know all files format to find a way for hiding information in those files. This technique is difficult because there are always large numbers of the file format and some of them have no way to hide information in them**.**

*B. System Concept*

Concept of this system can be summarized as hiding the password or any information beyond the end of an executable file so there is no function or routine (open-file, read, write, and close-file) in the operating system to extract it. This operation can be performed in two alternative methods:

- Building the file handling procedure independently of the operating system file handling routines. In this case we need canceling the existing file handling routines and developing a new function which can perform our need, with the same names. This way needs the customer to install the system application manually as shown in Figure 2.

- Developing the file handling functions depending on the existing file handling routines. This way can be performed remotely as shown in Figure 3.

The advantage of the first method is it doesn't need any additional functions, which can be identified by the analysts. The disadvantage of this method is it needs to be installed (can not be operated remotely). The advantage of the second method is it can be executed remotely and suitable for networks and the internet applications. So we choose this concept to implementation in this paper.

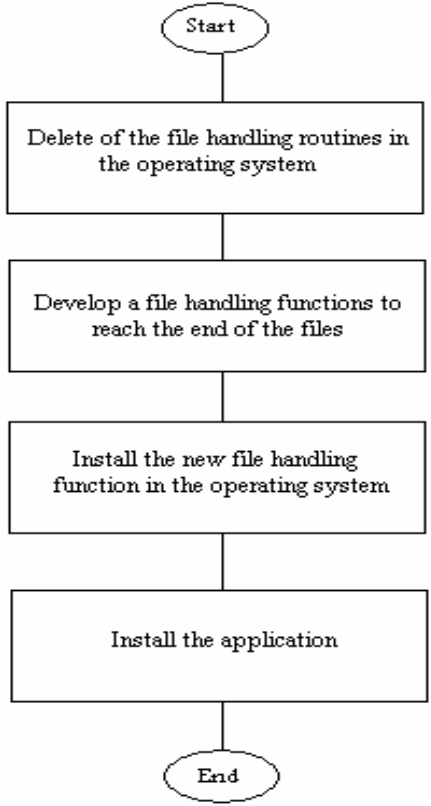

Figure 2. First Method of the System Concept

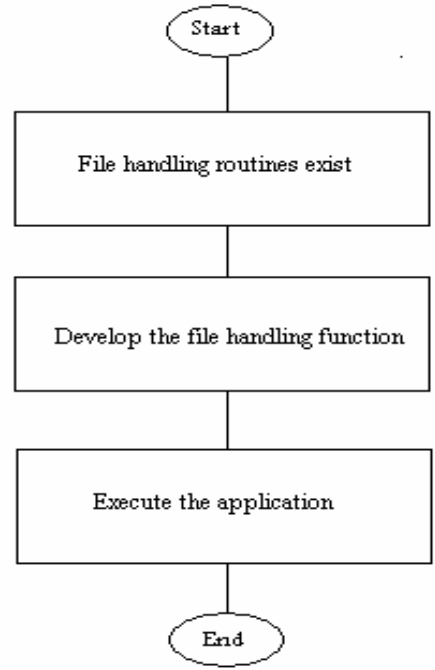

Figure 3. Second Method of the System Concept

C. *System Features*

This system has the following feature*:*
- The hiding operation within image page of EXE file using the statistical technique increases the degree of security for the information hiding which is used in the proposed system because the data which is embedded inside the EXE file is not embed directly of EXE file , it will be hiding within image page of EXE file. So the attacker can not be guessing the information hidden.

- The cover file can be executed normally after hiding operation. Because the hidden information already hide in the image page within exe.file and thus cannot be manipulated as the exe.file, therefore, the cover file still natural, working normally and not effected, such as if the cover is EXE file (WINDOWES XP SETUP) after hiding operation it'll continued working, In other words, the EXE file can be installed of windows.

- Virus detection programmers' can't detect such as files, the principle of antivirus check are checking from beginning to end. When checking the exe.files by antivirus, will checked it from beginning to end of it, since the principle of information hiding for this system within image page of EXE file .The information hiding will be hide inside the image page and the EXE file after hiding process is same manufacture of EXE file before hiding process. That is why the EXE file undetectable by Unit-Virus.

*D. The Proposed System Structure*

The system has been implementation by using Java. The block flow of hiding operation can be performed as shown in Figure 4. The following algorithm is the hiding operation procedure. The block flow of retract operation can be performed as shown in Figure 5. The following algorithm is the retract operation procedure.

**The following algorithm is the hiding operation procedure:**

Procedure: Hide operation.
Input: Hidden file name, cover file name.
Output: Stego-File.
- Begin (1).
- Opens the cover file (EXE file).
- Assign a pointer to the end of (Section header), which is before image page of the cover file.
- Select the image page consider normal page, it consider the cover for data.
    - Begin (2) for the image page.
    - Write the hidden file name.
    - Assign a pointer after hidden file name.
    - Write the hidden file content.
    - End(2) for the image page
- End (1).

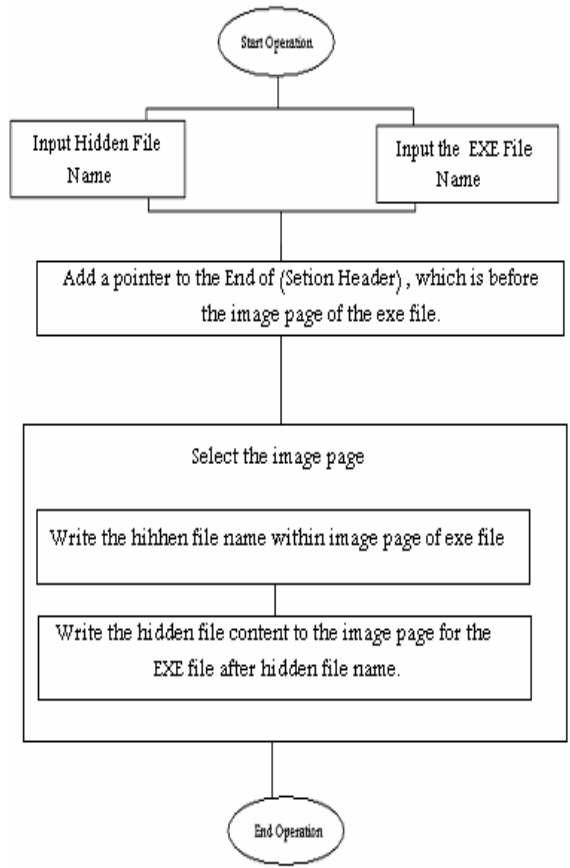

Figure 4. Block Flow of Hiding Operation.

**The following algorithm is the retract operation procedure:**
- Begin (1).
- Select the cover file.
- Get the End of the (Section Header) of EXE File.
- Select the image page:
    - Begin (2) for the image page.
    - Read the name of the hidden file.
    - Read the Hidden data.
    - Create a file using hiding file name.
    - Write in to the Create file the hiding data.
    - End (2) for the image page.
- End(1)

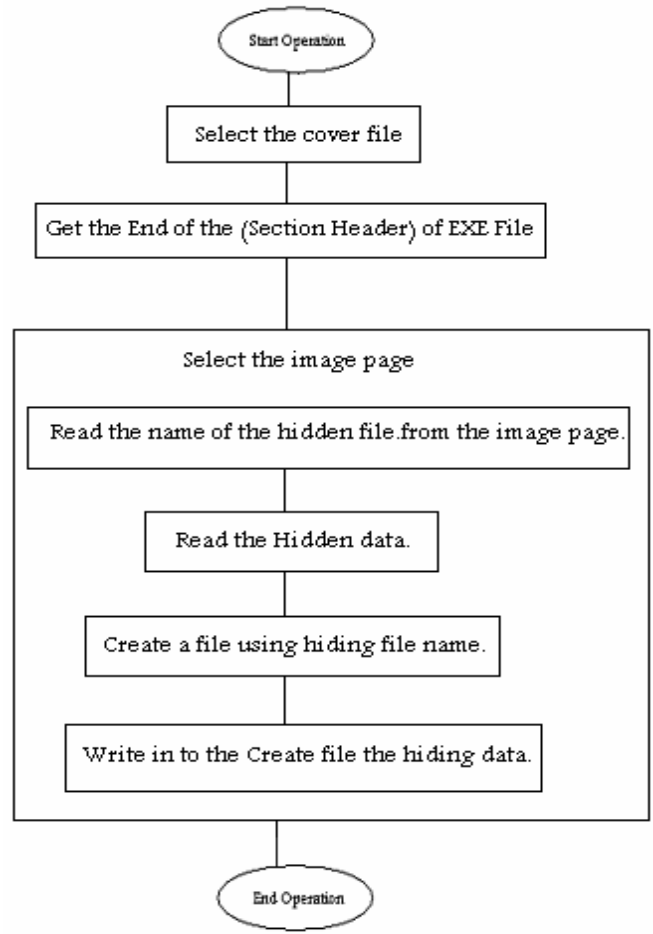

Figure 5. Block Flow of Retract Operation

V. CONCLUSION

The .EXE files are one of the most important files in operating systems and in most systems designed by developers (programmers/software engineers), and then hiding information in these file is the basic goal for this paper, because most users of any system cannot alter or modify the content of these files. We get the following conclusions:

- PE files structure is very complex because they depend on multi headers and addressing, and then insertion of data to PE files without full understanding of their structure may damage them, so the choice is to hide the information beyond the structure of these files, so the approach of the proposed system is to prevent the hidden information to observation of these systems.
- One of important conclusion most antivirus systems do not allow direct write in executable file, so the approach of the proposed system is to prevent the hidden information to observation of these systems.
- The cover file can be executed normally after hiding operation. Other word the cover file still natural, working normally and not affected.

## VI. FUTURE WORK

There are many suggestions for improving the proposed system, the main suggestions are:

- Developing the method which is used in proposed system to deal with other PE files such as "dll', "sys", "cpl", and "ocx".
- Developing the proposed system to deal with other executable files created by other operating systems like (LINUX, UNIX or OS/2).

## Author Information

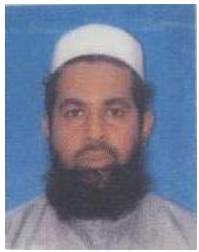

**Md. Rafiqul Islam** received his B Sc (Electrical and Electronic Engineering) from BUET, Dhaka in 1987. He received his MSc and PhD both in Electrical Engineering from UTM in 1996 and 2000 respectively. He is Fellow of IEB and member of IEEE. He is currently faculty member of Electrical and Computer Engineering Department of International Islamic University Malaysia. His area of research interest are radio link design, RF propagation measurement and RF design, smart antennas and array antennas design etc.

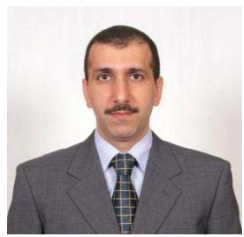

**Dr. Ahmed Wathik Naji** - He obtained his 1st Class Master degree in Computer Engineering from University Putra Malaysia followed by PhD in Communication Engineering also from University Putra Malaysia. He supervised many postgraduate students and led many funded research projects with more than 50 international papers. He has more than 10 years of industrial and educational experience. He is currently Senior Assistant Professor, Department of Electrical and Computer Engineering, International Islamic University Malaya, Kuala Lumpur, Malaysia.

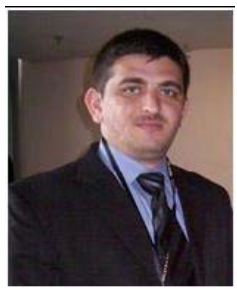

**Aos Alaa Zaidan** He obtained his 1st Class Bachelor degree in Computer Engineering from university of Technology / Baghdad followed by master in data communication and computer network from University of Malaya. He led or member for many funded research projects and He has published more than 40 papers at various international and national conferences and journals, currently he is working on the multi module for Steganography, Development & Implement a novel Skin Detector. He is members IAENG, WASET, and IACSIT.

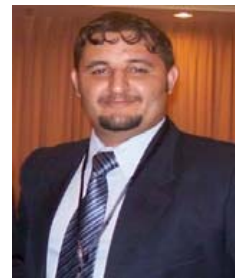

**Bilal Bahaa Zaidan** He obtained his bachelor degree in Mathematics and Computer Application from Saddam University/Baghdad followed by master from Department of Computer System & Technology Department Faculty of Computer Science and Information Technology/University of Malaya /Kuala Lumpur/Malaysia, He led or member for many funded research projects and He has published more than 40 papers at various international and national conferences and journals. He is members IAENG, WASET, and IACSIT.